\documentstyle[12pt]{article}
\newcommand{\be}{\begin{equation}}
\newcommand{\ee}{\end{equation}}
\newcommand{\bea}{\begin{eqnarray}}
\newcommand{\eea}{\end{eqnarray}}
\newcommand{\al}{\alpha}
\newcommand{\bt}{\beta}
\newcommand{\gm}{\gamma}

\newcommand{\dl}{\delta}
\newcommand{\Dl}{\Delta}
\newcommand{\eps}{\epsilon}
\newcommand{\zt}{\zeta}
\newcommand{\et}{\eta}

\newcommand{\kp}{\kappa}
\newcommand{\lm}{\lambda}
\newcommand{\Lm}{\Lambda}

\newcommand{\rh}{\rho}

\newcommand{\ta}{\tau}

\newcommand{\rarrow}{\rightarrow}
\newcommand{\nn}{\nonumber}

\begin{document}
\title{Higher-dimensional models in gravitational theories of quartic
Lagrangians}
\author{K. Kleidis, A. Kuiroukidis, D. B. Papadopoulos and 
H. Varvoglis\\
{\small Department of Physics}\\
{\small Section of Astrophysics, Astronomy and Mechanics}\\
{\small Aristotle University of Thessaloniki}\\
{\small 54006 Thessaloniki, GREECE} }

\maketitle

\begin{abstract}

Ten-dimensional models, arising from a gravitational action which 
includes terms up to the fourth order in curvature tensor, are 
discussed. The spacetime consists of one time direction and two 
maximally symmetric subspaces, filled with matter in the form of 
an anisotropic fluid. Numerical integration of the cosmological 
field equations indicates that exponential, as well as power-law, 
solutions are possible. We carry out a dynamical study of the 
results in the $H_{ext} - H_{int}$ plane and confirm the existence 
of {\em attractors} in the evolution of the Universe. Those 
attracting points correspond to {\em "extended" De Sitter} 
spacetimes in which the external space exhibits inflationary 
expansion, while the internal one contracts.\\

\vspace{.2in}


\end{abstract}

\section*{I. Introduction}

The mathematical background for a non-linear Lagrangian theory 
of gravity was first formulated by Lovelock$^{1}$, 
who proposed that the most general gravitational Lagrangian is
$$
{\cal L} = \sqrt {-g} \sum_{m=0}^{n/2} \lm_{(m)} {\cal L}_{(m)} 
\eqno (1.1)
$$
where $\lm_{(m)}$ are coupling constants, $n$ denotes the manifold's 
dimensions, $g$ is the determinant of the metric tensor and 
${\cal L}_{(m)}$ are functions of the Riemann curvature tensor, 
of the form
$$
{\cal L}_{(m)} = {1 \over 2^m} \dl_{\al_1 ... 
a_{2m}}^{\bt_1 ... \bt_{2m}} \; 
{\cal R}_{\bt_1 \bt_2}^{\al_1 \al_2} \: ... \: 
{\cal R}_{\bt_{2m-1} \bt_{2m}}^{\al_{2m-1} \al_{2m}} 
\eqno (1.2)
$$
where $\dl_{\bt}^{\al}$ is the Kronecker symbol, ${\cal L}_{(0)}$ is 
the volume $n$-form which gives rise to the cosmological constant, 
${\cal L}_{(1)} = {1 \over 2} {\cal R}$ is the Einstein-Hilbert (EH) 
Lagrangian and  ${\cal L}_{(2)}$ is the quadratic Gauss-Bonnett (GB) 
combination$^{2}$. Euler variation of the gravitational action 
corresponding to Eq.(1.1) yields the most general symmetric and 
divergenceless tensor, which describes the propagation of the 
gravitational field and depends only on the metric and its first and 
second order derivatives$^{1}$.

While quadratic Lagrangians have been widely studied (e.g. see 
Refs. [3,4] and references therein), cubic and/or quartic 
Lagrangians only recently have been introduced in the discussion of 
cosmological models in the framework of superstring 
theories$^{5-10}$. The reason is that, it is very hard to derive 
and (even harder) to solve the corresponding field equations. In 
this case, solutions may be obtained only through certain numerical 
techniques$^{11,12}$, where the idea of "{\em attractor}" plays a 
central role$^{13}$: If some special spacetime is the attractor for 
a wide range of initial conditions, such a spacetime is naturally 
realized asymptotically. Since the ten-dimensional superstring 
theory is a candidate for a realistic unified theory, it is very 
important to investigate whether a similar attractor exists in this 
theory.

In the present paper we integrate numerically the field equations, 
resulting from a quartic gravitational Lagrangian, to obtain 
anisotropic, ten-dimensional cosmological models. The spacetime 
consists of one time direction and two maximally symmetric 
subspaces, FRW$\otimes$FRW: The {\em external space}, representing 
the ordinary Universe and the {\em internal} one, constituted by 
the extra dimensions. The internal space is a compact manifold of 
very small {\em "physical size"} with respect to that of the 
{\em "visible"} space at the present epoch$^{14,15}$. Since, on the 
other hand, at the origin the two subspaces were of comparable 
physical size, the internal one must have somehow been contracted 
towards a static value of the order of Planck length, 
$l_{Pl} \sim 10^{-33} cm$, to achieve {\em "spontaneous 
compactification"}$^{16}$. Compactification is a topological process 
of quantum origin, which leads to the separation of the extra 
dimensions from the ordinary ones$^{17}$. In what follows 
we consider models of an already compactified internal space, 
i.e. we study only its contraction.

In Section II we derive the explicit form of the field equations for 
a quartic theory in ten dimensions, in which both subspaces are 
filled with an anisotropic fluid. In Section III we solve 
numerically the field equations, for a wide range of initial 
conditions and for several values of the "{\em free}" parameters 
involved, as regards {\bf (1)} vacuum models of flat subspaces 
and {\bf (2)} perfect fluid models of positively curved subspaces. 
Next, we carry out a dynamical study in the $H_{ext} - H_{int}$ 
plane, where each $H_j$ represents the Hubble parameter of the 
corresponding subspace. Accordingly, we confirm the existence 
of attracting points and investigate their evolution with respect 
to the variation of the coupling constants $\lm_{(m)}$. The 
explicit time-dependence of the unknown scale functions may be 
subsequently determined by solving the linearized field equations 
around these attracting points. The corresponding analysis is 
presented in Section IV.

\section*{II. The field equations in a quartic gravity theory}

We consider a ten-dimensional line element, representing 
cosmological models which consist of two homogeneous and isotropic 
factor spaces, of the form
$$
ds^2 \; = \; - \: dt^2 \; + \; R^2(t) \: {\sum_{i=1}^3 
\left ( dx^i \right )^2 \over 1 \: + \: {1 \over 4} k_{ext} 
\: \sum_{i=1}^3 \left ( x^i \right )^2} 
\; + \; S^2(t) \: {\sum_{j=4}^9 \left ( dx^j \right )^2 
\over 1 \: + \: {1 \over 4} k_{int} \: \sum_{j=4}^9 
\left ( x^j \right )^2}  
\eqno (2.1) 
$$
where $\hbar = 1 = c$, $R(t)$ and $S(t)$ are the cosmic scale 
functions of the external and the internal space respectively, 
$k_{ext} = -1, 0, +1$ is the curvature parameter of the 
{\em "ordinary"} space and $k_{int} = 0, +1$ is the corresponding 
parameter of the internal one. Therefore, the extra dimensions may 
be compactified either in a six-dimensional sphere, for 
$k_{int} = +1$, or in a six-dimensional torus, for $k_{int} = 0$. 
The spatial section of the metric (2.1) can be viewed as the direct 
product of two FRW models with three and six dimensions 
respectively$^{6}$. These models may be obtained through Hamilton's 
principle, from a ten-dimensional action in which the gravitational 
part is of the form 
$$
I \; = \; {1 \over V_{int}} \: \int \: \sqrt {-g} 
\left [ \lm_{(0)} {\cal L}_{(0)}
+ \lm_{(1)} {\cal L}_{(1)} + \lm_{(2)} {\cal L}_{(2)} 
+ \lm_{(3)} {\cal L}_{(3)} 
+ \lm_{(4)} {\cal L}_{(4)} \right ] \: d^{10} x 
\eqno (2.2)
$$
where each of ${\cal L}_{(m)}$ is given by Eq.(1.2), $\lm_{(m)}$ are 
the corresponding coupling constants and $V_{int}$ is a 
normalization constant$^{18}$, corresponding to the {\em "physical 
size volume"} of the internal space, once it may be considered 
static$^{3}$. The field equations read
$$
{\cal L}_{\mu \nu} \; = \; -\: 8 \pi G_{10} \: T_{\mu \nu} 
\eqno (2.3)
$$
where ${\cal L}_{\mu \nu}$ is the Lovelock tensor up to the fourth 
order in curvature (Greek indices refer to the ten-dimensional 
spacetime)$^{1}$ and $G_{10} = G \: V_{int}$ is the ten-dimensional 
gravitational constant$^{19}$. $T_{\mu \nu}$ is the energy-momentum 
tensor of an anisotropic perfect fluid source, of the form 
$T_{\mu \nu} = diag \: (\rh, -p_{ext},..., -p_{int},...)$, 
where $\rh$ is the total mass-energy density, while $p_{ext}$ and 
$p_{int}$ are the pressures associated to each factor space, 
separately. For the metric (2.1) Eq.(2.3) is decomposed into three 
independent equations of the form (cf. Ref. [9])
\bea
16 \pi G_{10} \: \rh &=& \lm_{(0)} + 6 \lm_{(1)} 
\left [ P + 5 Q + 6 ( {\dot{R} \over R } ) 
( {\dot{S} \over S } ) \right ] \nn \\
&+& 72 \lm_{(2)} \left [ 5 Q^2 + 5 P Q + 10 ( {\dot{R} \over R } )^2 
( {\dot{S} \over S } )^2 
+ 2 P ( {\dot{R} \over R } ) ( {\dot{S} \over S } ) 
+ 20 Q ( {\dot{R} \over R} ) 
( {\dot{S} \over S } ) \right ] \nn \\
&+& 720 \lm_{(3)} \left [ Q^3 + 8 ( {\dot{R} \over R} )^3 
( {\dot{S} \over S} )^3
+ 9 P Q^2 + 
18 Q^2 ( {\dot{R} \over R } ) ( {\dot{S} \over S } ) 
+ 12 P Q ( {\dot{R} \over R })
( {\dot{S} \over S } ) \right . \nn \\ 
&+& \left . 36 Q ( {\dot{R} \over R } )^2 
( {\dot{S} \over S } )^2 \right ] 
+ 17280 \lm_{(4)} \left [ P Q^3 \right . \nn 
\eea
$$
\left . + \; \; \; 6 Q^2 ( {\dot{R} \over R } )^2 
({\dot{S} \over S } )^2 + 
6 P Q^2 ( {\dot{R} \over R } ) ( {\dot{S} \over S } ) + 
8 Q ( {\dot{R} \over R } )^3 
( {\dot{S} \over S } )^3 \right ] \; \; \; \; 
\eqno (2.4a)
$$
\bea
-16 \pi G_{10} \: p_{ext} &=& \lm_{(0)} + 2 \lm_{(1)} 
\left [ P + 15 Q + 
12 ( {\dot{R} \over R } ) ({\dot{S} \over S }) +
2 ( {\ddot{R} \over R } ) 
+ 6 ( {\ddot{S} \over S } ) \right ] \nn \\ 
&+& 24 \lm_{(2)} \left [ 15 Q^2 + 10 ( {\dot{R} \over R } )^2 
( {\dot{S} \over S } )^2 +
5 P Q + 40 Q ( {\dot{R} \over R } ) ( {\dot{S} \over S } ) + 
20 Q ( {\ddot{S} \over S } ) \right . \nn \\
&+& \left . 2 P ( {\ddot{S} \over S } ) 
+ 10 Q ( {\ddot{R} \over R } ) + 20 ( {\ddot{S} \over S } )
( {\dot{R} \over R } ) ( {\dot{S} \over S } ) 
+ 4 ( {\ddot{R} \over R } )
( {\dot{R} \over R } ) ( {\dot{S} \over S } ) \right ]  \nn \\
&+& 720 \lm_{(3)} \left [ Q^3 + 3 Q^2 P 
+ 12 Q^2 ( {\dot{R} \over R } ) 
( {\dot{S} \over S } ) + 6 Q^2 ( {\ddot{S} \over S } ) 
+ 6 Q^2 ({\ddot{R} \over R } ) \right . \nn \\ 
&+& \left . 12 Q ( {\dot{R} \over R } )^2 ( {\dot{S} \over S } ) 
+ 4 P Q ( {\ddot{S} \over S } ) + 8 ( {\ddot{S} \over S } )
( {\dot{R} \over R } )^2 ( {\dot{S} \over S } )^2
+ 24 Q ( {\ddot{S} \over S } ) ( {\dot{R} \over R } ) 
( {\dot{S} \over S } ) 
\right . \nn \\
&+& \left . 8 Q ( {\ddot{R} \over R } ) ( {\dot{R} \over R } ) 
( {\dot{S} \over S } ) \right ]  + 5760 \lm_{(4)} \left [ 2 Q^3 
( {\ddot{R} \over R } ) \right . \nn \\
&+& 12 Q^2 ( {\ddot{S} \over S } ) ( {\dot{R} \over R } ) 
( {\dot{S} \over S } ) + 12 Q^2 ( {\ddot{R} \over R } ) 
( {\dot{R} \over R } ) 
( {\dot{S} \over S } ) + 6 P Q^2 ( {\ddot{S} \over S } )  \nn 
\eea
$$
+ \; \; \left . 24 Q ( {\ddot{S} \over S } ) 
( {\dot{R} \over R } )^2 ( {\dot{S} \over S } )^2 +
P Q^3 + 6 Q^2 ( {\dot{R} \over R } )^2 
( {\dot{S} \over S } )^2 \right ] \; \; 
\eqno (2.4b)
$$
\bea
-16 \pi G_{10} \: p_{int} &=& \lm_{(0)} 
+ 2 \lm_{(1)} \left [ 3 P + 10 Q + 15 ( {{\dot R} \over R} ) 
( {{\dot S} \over S} ) + 3 ( {{\ddot R} \over R} )
+ 5 ( {{\ddot S} \over S} ) \right ] \nn \\
&+& 24 \lm_{(2)} \left [ 5 Q^2 + 20 ( {{\dot R} \over R} )^2 
( {{\dot S} \over S} )^2 
+ 10 P Q + 5 P ( {{\dot R} \over R} ) ( {{\dot S} \over S} ) + 
P ( {{\ddot R} \over R} ) \right . \nn \\
&+& \left . 30 Q ( {{\dot R} \over R} ) ( {{\dot S} \over S} ) 
+ 10 Q ( {{\ddot R} \over R} ) + 5 P ( {{\ddot S} \over S} ) 
+ 10 Q ( {{\ddot S} \over S} ) + 10 ( {{\ddot R} \over R} ) 
( {{\dot R} \over R}) 
( {{\dot S} \over S} ) \right . \nn \\
&+& \left . 20 ( {{\ddot S} \over S} ) ( {{\dot R} \over R} ) 
( {{\dot S} \over S} ) \right ] \nn \\
&+& 720 \lm_{(3)} \left [ 4 ( {{\dot R} \over R} )^3 
( {{\dot S} \over S} )^3 
+ 3 P Q^2 + 3 Q^2 ( {{\dot R} \over R} ) ( {{\dot S} \over S} ) 
+ 3 Q^2 ( {{\ddot R} \over R} ) \right . \nn \\
&+& \left . 2 Q^2 ( {{\ddot S} \over S} ) 
+ 6 P Q ( {{\dot R} \over R} ) ( {{\dot S} \over S} ) 
+ 12 Q ( {{\dot R} \over R} )^2 ( {{\dot S} \over S} )^2 
+ 2 P Q ( {{\ddot R} \over R} ) \right . \nn \\
&+& \left . 6 P Q ( {{\ddot S} \over S} ) 
+ 4 ( {{\ddot R} \over R} ) ( {{\dot R} \over R} )^2 
( {{\dot S} \over S} )^2
+ 12 ( {{\ddot S} \over S} ) ( {{\dot R} \over R} )^2 
( {{\dot S} \over S} )^2 
+ 4 P ( {{\ddot S} \over S} ) ( {{\dot R} \over R} ) 
( {{\dot S} \over S} ) 
\right . \nn \\
&+& \left . 12 Q ( {{\ddot R} \over R} ) ( {{\dot R} \over R} ) 
( {{\dot S} \over S } )
+ 12 Q ( {{\ddot S} \over S} ) ( {{\dot R} \over R} ) 
( {{\dot S} \over S} ) \right ] \nn \\
&+& 5760 \lm_{(4)} \left [ 6 Q^2 ( {{\ddot R} \over R} ) 
( {{\dot R} \over R} ) 
( {{\dot S} \over S} ) + 3 P Q^2 ( {{\ddot S} \over S} ) 
+ 12 Q ( {{\ddot S} \over S} ) ( {{\dot R} \over R} )^2 
( {{\dot S} \over S} )^2 \right . \nn \\
&+& \left . 12 P Q ( {{\ddot S} \over S} ) ( {{\dot R} \over R} ) 
( {{\dot S} \over S} ) 
+ 3 P Q^2 ( {{\dot R} \over R} ) ( {{\dot S} \over S} ) 
+ 4 Q ( {{\dot R} \over R} )^3 ( {{\dot S} \over S} )^3 \right . \nn 
\eea
$$
+ \left . 3 P Q^2 ( {{\ddot R} \over R} ) 
+ 12 Q ( {{\ddot R} \over R} ) ( {{\dot R} \over R} )^2 
( {{\dot S} \over S} )^2 
+ 8 ( {{\ddot S} \over S} ) ( {{\dot R} \over R} )^3 
( {{\dot S} \over S} )^3 \right ] 
\eqno (2.4c) 
$$
where an overdot denotes derivative with respect to time and we have 
set
$$
P = ( {{\dot R} \over R} )^2 + {k_{ext} \over R^2} \; \; , \; \;  
Q = ( {{\dot S} \over S} )^2 + {k_{int} \over S^2} 
\eqno (2.5)
$$
Since the Lovelock tensor is divergenceless, 
${\cal L}_{ \; \; ; \nu}^{\mu \nu} = 0$,
we obtain the conservation law $T_{ \; \; ; \nu}^{\mu \nu} = 0$, 
which gives
$$
{\dot \rh} \; + \; 3 \left ( \rh \: + \: p_{ext} \right ) 
{{\dot R} \over R} \; 
+ \; 6 \left ( \rh \: + \: p_{int} \right ) 
{{\dot S} \over S} = 0 
\eqno (2.6)
$$
Further inspection of the system of Eqs. (2.4) and (2.6) shows that 
only three of them are truly independent. Thus, the problem is 
completely determined by those, plus the two equations of state for 
the matter content, one for each subspace$^{17}$. In the present 
article we consider two cases with regard to the energy-momentum 
tensor: {\bf (a)} Vacuum models, $\rh = 0$, in connection to flat 
spatial sections $( k_{ext} = 0 = k_{int} )$ and {\bf (b)} Models 
of an heterotic superstring gas$^{20}$, $p_{ext} = {1 \over 3} \rh $ 
and $p_{int} = 0$, in connection to possitively curved spatial 
sections $(k_{ext} = 1 = k_{int})$. In the later case, the 
conservation law (2.6) gives
$$
\rh \; = \; {M \over R^4 \: S^6} \eqno (2.7)
$$
where M is an integration constant. Thus, the external space is 
radiation dominated$^{9,11,20}$.

In principle, we may integrate the system of Eqs. (2.4) and (2.6) 
to obtain the form of the unknown scale functions. However this is 
not an easy task, even in the most simple and symmetric cases$^{9}$.
Nevertheless, we may get a good estimation of their dynamic behaviour 
through numerical integration$^{11}$.

Once the two equations of state are determined, Eq.(2.6) may be 
readily solved to give the unknown energy density and pressures, 
as functions of $R(t)$ and $S(t)$. These expressions are subsequently 
introduced in the r.h.s. of Eqs.(2.4). Now, only two of these 
equations are truly independent. The third one corresponds to a 
{\em constraint}, to be satisfied by the solutions of the system. 
As such, we choose Eq.(2.4$a$). The remaining
independent field equations (2.4$b$) and (2.4$c$) may be recast in
the form of a first order system (see also [11]), as follows
$$
{\dot H}_{ext} = G_1 \: \left ( H_{ext}\, , \; H_{int} \, , 
\; X \, , \; Y \right ) 
\eqno (2.8a)
$$
$$
{\dot H}_{int} = G_2 \: \left ( H_{ext}\, , 
\; H_{int} \, , \; X \, , \; Y \right ) 
\eqno (2.8b)
$$
$$
{\dot X} = - \: X \: H_{ext} \eqno (2.8c)
$$
$$
{\dot Y} = - \: Y \: H_{int} \eqno (2.8d)
$$
where we have set
$$
H_{ext} = {{\dot R} \over R} \; , \; \; H_{int} = {{\dot S} \over S} 
\; , \; \; X^2  =  {k_{ext} \over R^2} \; , \; \; Y^2  =  {k_{int} 
\over S^2} 
\eqno (2.9)
$$
and the explicit forms of the functions $G_1$ and $G_2$ are given 
in the Appendix A.

Finally, it is convenient to make a parameter rescaling in the 
field equations, of the form
$$
\kp_m \; = \; {\lm_{(m)} \over \lm_{(1)}} \; , \; \; m \: 
= \: 0, \, 1, \, 2, \, 3, \, 4
\eqno (2.10)
$$
where $\lm_{(1)} = (16 \pi G)^{-1}$ is the coupling constant in the 
four-dimensional General Relativity (GR). The value of the 
normalized coupling constants, $\kp_m$ $(\kp_m \leq 1)$, is directly 
proportional to the contribution in the field equations of the 
corresponding $m-th$ order non-linear term, with respect to the 
results obtained in the EH cosmology. Clearly, $\kp_1 = 1$.

\section*{III. Numerical Results}

We integrate numerically the system of Eqs.(2.8). The constraint 
(2.4$a$) is checked to be satisfied with an accuracy of $10^{-10}$ 
along integration. The initial conditions $H_0^{ext}, \; 
H_0^{int}, \; X_0, \; Y_0$ are chosen so that: {\bf (a)} $X_0 
= Y_0$, i.e. at the origin, the two factor spaces are separated, 
but of the same "physical size"$^{16}$. {\bf (b)}
$H_0^{ext} > 0$, i.e. initially the ordinary space expands, in 
accordance to what we observe at the present epoch$^{12,21}$. 
{\bf (c)} $H_0^{int} < 0$, i.e. at the origin, the internal space 
contracts, in correspondance to "spontaneous compactification" 
$^{12,16,17,21}$. The cases where either $H_0^{ext} < 0$ or 
$H_0^{int} > 0$ are not permitted, since the constraint equation 
is not satisfied. Nevertheless, the case where both conditions 
$H_0^{ext} < 0$ and $H_0^{int} > 0$ are valid is acceptable by 
numerical analysis. Actually, it corresponds to the 
time-reversed solution of the system (2.8).

The time coordinate is measured in dimensionless units, being 
normalized with respect to the Planck time, $\ta = t/t_{Pl}$ 
($t_{Pl} = \sqrt {G} \sim 10^{-43} sec$). The limits of numerical 
integration range from $\ta = 0$ to $\ta = 10^5$. The upper limit 
coincides with the origin of the GUT epoch$^{21}$, $t_{GUT} = 10^5 
\: t_{Pl}$, corresponding to the end of the string regime$^{22}$. 
However we have to point out that, although the origin of the time 
coordinate is set at $\ta = 0$, the equations (2.8) may not be valid 
in the region $0 < \ta \leq 1$ since, in the absence of a quantum 
gravity theory, there is always a region of ambiguity around 
$t = 0$, of the order of Planck time$^{23-25}$.

The solution of the system (2.8) may be represented as curves in the
$H_{ext}-H_{int}$ plane. Any point located on these curves always 
satisfies the constraint condition (2.4$a$). Thus, the curves 
actually represent {\em "orbits"} of the dynamical system under 
study. Each curve, corresponding to a different set of initial 
conditions, is bounded by fixed points (or infinities) and 
represents a different type of evolution for the Universe. 

In what follows, we focus attention on the existence and the 
evolution of attracting points in the $H_{ext}-H_{int}$ plane. 
The reason rests in the physical meaning of the {\em attractor:} 
No matter what the behaviour of a cosmological model at the origin 
might be, it will always end up to evolve as indicated by the 
location of the attracting point in the $H_{ext}-H_{int}$ plane.

\subsection*{(1) Vacuum models with spatially flat subspaces}

We study the evolution of vacuum ten-dimensional cosmological 
models, with metric of the form (2.1), in which both subspaces 
are spatially flat, i.e. $k_{ext} = 0 = k_{int}$. Thus, $X=0=Y$. 

The first case to study are the GB models (see also Refs. [11,12]). 
In this case, $\kp_2 = 1$ and $\kp_0 = 0 = \kp_3 = \kp_4$. The 
non-linear curvature contributions to the field equations come out 
from the quadratic terms alone. The time evolution of the Hubble 
parameters is presented in Fig. 1a. We see that both parameters 
evolve to approach constant values in the later stages. This 
situation verifies the existence of attracting points in the 
$H_{ext}-H_{int}$ plane during the evolution of the Universe. 
Therefore, for a wide range of initial conditions, both subspaces 
will end up to evolve as De Sitter spaces, in complete 
correspondence to the results of Ishihara$^{12}$. 

We also observe that $H_{ext} > 0$ and $H_{int} < 0$. Therefore, 
while the internal space contracts exponentially to achieve 
spontaneous compactification, the external one expands, a fact that 
corresponds to an inflationary phase. This result indicates that the 
introduction of the non-linear curvature terms into the 
gravitational action may play an important role as far as the 
inflation is conserned$^{26-30}$. The explicit location of the 
attracting point is shown in Fig. 1b. The attractor corresponds 
to the fixed point $D_2$ recognized by Ishihara$^{12}$ in the 
evolution of the extended De Sitter models in GB theory.

The next step is to introduce into the problem a {\em "bare"} 
cosmological constant, $\Lm$, corresponding to the expectation 
value of the vacuum energy density$^{25}$. Now, in addition to 
$\kp_2$, we also have $\kp_0 \neq 0$, while $\kp_3 = 0 = \kp_4$. 
When $\kp_0 \: \in \: [0 \: , \: 1]$ the value of the cosmological 
constant in physical units is $\Lm = 2 \kp_0 \times 10^{-48} 
\: cm^{-2}$, which is quite small. 

The behaviour of the model is qualitatively similar to the previous 
case. Again we verify the existence of an "attractor". Both 
subspaces correspond to De Sitter models. The external space 
exhibits inflationary expansion, while the internal one contracts. 
However, in this case, the location of the attracting point $D_2$ has 
changed to higher absolute values in the evolution of $H_{ext} $ and 
$H_{int}$ (Fig. 2a). We may determine explicitly the law of the 
attractor's displacement in the $H_{ext}-H_{int}$ plane, caused by 
variations of the cosmological constant.

In general, to determine the exact location of the attracting points 
in an $H_{ext}-H_{int}$ plane, requires to set
$$
G_1(H_{ext}, H_{int}, X, Y) = 0 \eqno (3.1.1a)
$$
$$
G_2(H_{ext}, H_{int}, X, Y) = 0 \eqno (3.1.1b)
$$
In the case of flat and vacuum subspaces $(X = Y = p_{ext} 
= p_{int} = 0)$, Eqs.(3.1.1) read
$$
f_1 \left ( H_{ext}, H_{int}, \kp_m \right ) 
= \left [ G_{12}G_{20}-G_{22}G_{10} 
\right ]_{X=Y=0} \; = \; 0 
\eqno (3.1.2a)
$$
$$
f_2 \left ( H_{ext}, H_{int}, \kp_m \right ) 
= \left [ G_{21}G_{10}-G_{11}G_{20} 
\right ]_{X=Y=0} \; = \; 0 
\eqno (3.1.2b)
$$
where $m = 0, 2, 3, 4$ and the quantities $G_{ij}$ are presented 
in the Appendix A. We differentiate the functions $f_1$ and $f_2$ 
with respect to $H_{ext} \: , \: H_{int}$ and $\kp_m$, to obtain 
a system of first order differential equations ({\em "variational 
equations"}) 
$$
df_1 = ({\partial f_1 \over \partial H_{ext} })_P dH_{ext} 
+ ({\partial f_1 \over \partial H_{int} })_P dH_{int} 
+ \sum_j ({\partial f_1 \over \partial \kp_m })_P d \kp_m = 0 
\eqno (3.1.3a)
$$
$$
df_2 = ({\partial f_2 \over \partial H_{ext} })_P dH_{ext}
+ ({\partial f_2 \over \partial H_{int} })_P dH_{int}
+ \sum_j ({\partial f_2 \over \partial \kp_m })_P d \kp_m =0 
\eqno (3.1.3b)
$$
The system (3.1.3) may be used, to determine the evolution of the 
attracting point $D_2 \; (H_{ext}, H_{int})$, under the variation 
of the normalized coupling constants $\kp_m$. For $\kp_2 = 1$, in 
the case of vanishing $\kp_3$ and $\kp_4$, the evolution of the 
attractor $D_2 \; (H_{ext},H_{int})$ with respect 
to the variation of the cosmological constant $\kp_0$, is given by
$$ 
({dH_{ext} \over d \kp_0 }) = ({QQ_1 \over PP}) \eqno (3.1.4a)
$$
$$
({dH_{int} \over d \kp_0 }) = ({QQ_2 \over PP}) \eqno (3.1.4b)
$$
where the functions $PP$, $QQ_1$ and $QQ_2$ are given in the 
Appendix B. Subsequently, the system (3.1.4) is evaluated by 
numerical integration. The corresponding results are shown in 
Fig. 2b. Using least square fitting, we see that the displacement 
of $D_2$ takes place along the straight line
$$ 
H_{int} = -0.075 H_{ext} -0.071 \eqno (3.1.5)
$$

The investigation of the behaviour of the models under consideration 
by including a third order curvature term, corresponds to study them 
at earlier epochs in the history of the Universe. Indeed, if we are 
interested in the behaviour of the model very close to the initial 
singularity, the leading terms to consider in the field equations 
are those with the highest power in $\left ({1 \over t} \right )$, 
i.e. those obtained from the highest order terms in the 
gravitational action$^{9}$.

The time-evolution of the model is quite similar to the previous 
cases. In the later stages it corresponds to an extended De Sitter 
model, in which both subspaces evolve exponentially. The external 
space expands, while the internal one contracts (Fig. 3a).

Again, we verify the existence of an attracting point P in the 
evolution of the Hubble parameters and we investigate its behaviour 
as $\kp_3$ increases, from 0 to 1, i.e. until it becomes as 
important as the quadratic term. The evolution of the attractor in 
the ${\cal L}_{(3)}$-theory, with respect to the variation of 
$\kp_3$, may be obtained in a similar way as in the $\kp_0$ case. 
We differentiate the functions $f_1$ and $f_2$ with respect to 
$H_{ext} \: , \: H_{int}$ and $\kp_3$ to obtain a first order system 
of differential equations which, for $\kp_2 = 1$ and for vanishing 
$\kp_0$ and $\kp_4$, will determine the displacement of P in the 
$H_{ext}-H_{int}$ plane, under the variation of $\kp_3$. 

The corresponding results are presented in Fig. 3b. We observe 
that the attractor moves to higher absolute values of $H_{int}$ 
as $\kp_3$ increases. This result has a clear physical meaning. 
Since increasing $\kp_3$ corresponds to study the earlier stages 
in the evolution of the Universe, we see that at these epochs 
the internal space contracts at higher rates than those of the 
GB theory. Then Fig. 3b verifies that at the late stages, where 
the GB theory holds alone, the value of the internal Hubble 
parameter decreases in order to achieve stabilization.

Again, the law of displacement of P in the $H_{ext}-H_{int}$ plane 
may be estimated using {\em best-fit methods}. In this context, we 
find that it may be represented by a sixth-order polynomial $H_{ext} 
= p_6 (H_{int})$, with coefficients: $ a_0 = 0.7373 \: , 
a_1 = -25.594 \: , \: a_2 = -457.324 \: , \: a_3 = -3323.9 \: , \: 
a_4 = -12156.9 \: , \: a_5 = -22135.2 $ and $a_6 = - 15905.6$.

Finally, to solve the cosmological field equations when all terms in 
the action (2.2) are included (i.e. $\kp_4 \neq 0 $), corresponds to 
study the dynamic behaviour of the model under consideration at even 
earlier epochs. The results are slightly different from those of the 
previous case (Fig. 4a). Again, in the later stages, the model 
consists of two De Sitter subspaces and there exists an attracting 
point. The attractor's displacement in the $H_{ext}-H_{int}$ plane 
is obtained in a way similar to the $\kp_0$ and $\kp_3$ cases and 
may be represented by a third-order polynomial, $H_{int} = 
p_3(H_{ext})$, with coefficients: $b_0 = 7.47 \: , \: b_1 = -34.27 
\: , \: b_2 = 51.33 $ and $b_3 = -25.74$. The corresponding result 
is shown in Fig. 4b.

Hence, we may conclude that in every case where non linear terms are 
included, the {\em "extended"} De Sitter solution (i.e. an 
exponentially expanding external space in connection to an 
exponentially contracting internal one) corresponds to an 
{\em "attractor"} of the dynamical system under consideration. 
Accordingly, (in our model) no matter how the Universe may 
originate, there is at least one period during its time-evolution 
in which it exhibits inflation of the ordinary space, accompanied 
by spontaneous compactification of the internal one$^{12,30}$.

\subsection*{(2) Perfect fluid models of curved subspaces}

We consider a ten-dimensional metric of the form (2.1), which now 
represents a class of cosmological models with positively curved 
subspaces $(k_{ext} = 1 = k_{int})$. Then, $X = R^{-1}(t)$ and 
$Y = S^{-1}(t)$ and we study the time-evolution of the cosmological 
models as results from the solution of the system (2.8). 

The numerical analysis is carried out in the same fashion as in the 
previous case of vacuum models. We consider that at the origin both 
subspaces are of the same "physical size", i.e. $X_0 = Y_0$, but 
they have different expansion rates, $H_0^{ext}$ and $H_0^{int}$. 
As such we choose the corresponding range used in the vacuum case. 
We normalize both scale functions $R(t)$ and $S(t)$ to unity, with 
respect to their value at the Planck epoch. That is 
$$
R(t) \rarrow {R(t) \over R_{Pl}} \; , \; \; S(t) \rarrow 
{S(t) \over S_{Pl}} 
\eqno (3.2.1)
$$
where $R_{Pl} \; = \; S_{Pl}$. As initial conditions we choose 
$R_0 = 100 = S_0$.

We represent the matter filling the Universe by a closed or 
heterotic superstring perfect gas, with the following equation 
of state, deduced by Matsuo$^{20}$
$$
p_{ext} \: = \: {1 \over 3} \: \rh \; , \; \; \; 
p_{int} \: = \: 0 
\eqno (3.2.2)
$$
Thus, the external space is radiation-dominated, while the internal 
one is pressureless. It has been recently shown that, in this case, 
the two subspaces are completely disjoint$^{4,31}$. The 
time-evolution of the total mass-energy density $\rh$ is accordingly 
given by Eq.(2.7).

As regards the GB models $(\kp_3 = 0 = \kp_4)$, we have performed a 
number of computational runs, varying the initial values of the 
Hubble parameters and the coupling constant $\kp_2$ as well, from 
$\kp_2 = 0.1$ to $\kp_2 = 1$. Numerical results in this case 
indicate that there is a considerable difference with respect to 
the vacuum-flat models. It rests in the fact that the range of 
values of the coupling constant $\kp_2$ may be splitted 
into two parts. Each one of these parts leads to a different 
time-evolution of both the external and the internal scale functions.

The first part consists of values of $\kp_2$ in the interval $0.1 
\leq \kp_2 \leq 0.65$, i.e. when the contribution of the quadratic 
curvature terms is relatively small. In this case we expect that the 
time-evolution of the Universe will be only slightly different from 
the corresponding EH one. Indeed, the numerical results indicate 
that the system (2.8) admits solutions with a power law dependence 
of the scale functions upon time, of the form
$$
R(t) \propto t^{m_1} \eqno (3.2.3a)
$$
$$
S(t) \propto t^{-m_2} \eqno (3.2.3b)
$$
where the values of the indices $m_1$ and $m_2$ are continuously 
increasing in the ranges $0.25 \leq m_1 \leq 0.55$ and $0.01 
\leq m_2 \leq 0.11$, as $\kp_2$ increases from $0.1$ to $0.65$. 
In this case, there are no attracting points in the evolution of 
the Universe. The last values in those ranges ($0.55$ and $0.11$, 
respectively), both corresponding to the value $\kp_2 = 0.65$, 
represent a Kasner-type regime$^{12,32-34}$ of the GB models. 
Indeed, the analytic approach in this case suggests that the two 
subspaces evolve as 
$$
R(t) \sim t^{p_1} \; , \; \; S(t) \sim t^{-p_2} \eqno (3.2.4)
$$
where both $p_1$ and $p_2$ are possitive and in a ten-dimesional 
spacetime they satisfy the conditions
$$
3 p_1 - 6 p_2 = 1 \eqno (3.2.5a)
$$
$$
3 p_1^2 + 6 p_2^2 = 1 \eqno (3.2.5b)
$$
The only physically acceptable solution of the system (3.2.5), 
compatible with the condition $p_1 \:, \: p_2 \; > \; 0$, is
$$
p_1 = {5 \over 9} = 0.555 \; , \; \; p_2 = {1 \over 9} = 0.111
$$
Therefore, when $\kp_2 = 0.65$, although the spatial sections are 
curved, the time-evolution of the Universe admits a Kasner-type 
solution. This solution actually lies on the interface between 
two different types of cosmological behaviour (Figs. 5a and 5b).

The second type of time-evolution arises when $ 0.65 < \kp_2 
\leq 1$. Then the Universe behaves, again, as an {\em extended} 
De Sitter spacetime (where the external space expands while the 
internal one contracts, both exponentially). In this case there 
exists an attracting point as in the vacuum-flat models (Fig. 6a). 

In conclusion, for a curved ten-dimensional GB cosmological model, 
filled with matter in the form of a superstring perfect gas, we 
may obtain three different types of cosmological behaviour, 
depending on the exact value of the normalized coupling constant 
$\kp_2$: \\

{\bf (a)} Power-law solutions, with no attracting points, 
when $0.1 \leq \kp_2 < 0.65$. \\

{\bf (b)} A Kasner-type model, when $\kp_2 = 0.65$. \\

{\bf (c)} Extended De Sitter models, with an attracting point, 
when $0.65 < \kp_2 \leq 1$. \\

In all cases, the external space expands, while the internal one 
contracts. The inclusion of the contribution of the third and/or 
the fourth order terms in the field equations simply amounts to a 
modulation of those results (Fig. 6b).

\section*{IV. Analytic Results}

Analytic expressions, for the time-evolution of the model Universe 
considered, may be obtained by solving the cosmological field 
equations (2.8) around the attracting points. Accordingly, we 
investigate the cosmological behaviour of a vacuum, ten-dimensional 
model with spatially flat subspaces $(X = 0 = Y)$ within the context 
of the quartic Lagrangian theory under consideration. Clearly, 
setting some of the coupling constants $\lm_{(m)}$ equal to zero 
corresponds to reducing the general theory to its lower case 
counterparts (EH-cosmology, GB-theory etc.).

Since we are interested in the behaviour of the model around the 
attracting points, we consider the linearized equations
$$
H_{ext} = A_1 + H_1(t) \eqno (4.1a)
$$
$$
H_{int} = A_2 + H_2(t) \eqno (4.1b)
$$
where $A_1$ and $A_2$ are the coordinates of the attractor, while 
$H_1(t)$ and $H_2(t)$ represent small perturbations around those 
values $( \vert H_1 \vert \:, \: \vert H_2 \vert \; \ll \: 1)$. 
Therefore, to obtain the time-evolution of $H_{ext}$ and $H_{int}$, 
we only have to solve the system (2.8) linearized with respect to 
$H_1$ and $H_2$.

The system of the cosmological field equations (2.8), linearized 
with respect to $H_1$ and $H_2$, may be written in the form
$$
{\dot H}_1(t) = {\bt_1 H_1 + \bt_2 H_2 + \bt_3 \over \al_1 H_1 
+ \al_2 H_2 + \al_3} 
\eqno (4.2a)
$$
$$
{\dot H}_2(t) = {\gm_1 H_1 + \gm_2 H_2 + \gm_3 \over \al_1 H_1 
+ \al_2 H_2 + \al_3} 
\eqno (4.2b)
$$
where $\al_j, \bt_j$ and $\gm_j$ $(j = 1,2,3)$ are constants, 
calculated directly from the linearization of the original 
equations, which depend on $A_1 \: , \: A_2$ and $\lm_{(m)} 
\; (m = 0,1,2,3,4)$. From Eqs.(4.2) we obtain
$$
{d H_1 \over d H_2} = {\bt_1 H_1 + \bt_2 H_2 
+ \bt_3 \over \gm_1 H_1 + \gm_2 H_2 + \gm_3 } 
\eqno (4.3)
$$
The solution of Eq.(4.3), in connection to Eqs.(4.1), will give, 
in the linear approximation, the analytic expression of $H_{ext}$ in 
terms of $H_{int}$. To solve Eq.(4.3), we need to have the solution 
$(h_1 \: , \: h_2)$ of the algebraic 
system
$$
\bt_1 H_1 + \bt_2 H_2 + \bt_3 = 0 \eqno (4.4a)
$$
$$
\gm_1 H_1 + \gm_2 H_2 + \gm_3 = 0 \eqno (4.4b)
$$
We choose
$$
\gm_1 \neq  0 \; \; , \; \; \bt_2 \gm_1 - \bt_1 \gm_2 \neq 0 
\eqno (4.5)
$$
and furthermore, we set
$$
w = H_1 - h_1 \eqno (4.6a)
$$
$$
z = H_2 - h_2 \eqno (4.6b)
$$
We verify that the solution of Eq.(4.3) depends on several algebraic 
combinations of the constants $\al_j \: , \: \bt_j$ and $\gm_j$, 
something that leads to several conditions between the coupling 
constants $\lm_{(m)}$. Therefore, we consider the following cases: \\

{\bf (a)} $\bt_1 + \gm_2 \neq 0$: This combination corresponds to 
the most general case. Setting
$$
\Dl = - \left [ 4 \bt_2 \gm_1 
+ \left ( \bt_1 - \gm_2 \right )^2 \right ] 
\eqno (4.7)
$$
the solution of Eq.(4.3) reads$^{35}$ 
$$
\ln { {1 \over c} \left [ \bt_2 z^2 + (\bt_1 - \gm_2) z w 
- \gm_1 w^2 \right ]} \; = \; 
\left \{ \begin{array}{ll}
{\bt_1 + \gm_2 \over \sqrt {-\Dl}} \ln {{\bt_1 - \gm_2 
- \sqrt {- \Dl} - 2 \gm_1 
{w \over z} \over \bt_1 - \gm_2 + \sqrt {- \Dl} 
- 2 \gm_1 {w \over z} }} & 
\mbox{for $\Dl < 0$} \\ \\
- {2 (\bt_1 + \gm_2) \over (\bt_1 -\gm_2) 
- 2 \gm_1 {w \over z} } & 
\mbox{for $\Dl = 0$} \\ \\
2 {\bt_1 + \gm_2 \over \sqrt {\Dl} } 
\arctan {{ (\bt_1 -\gm_2) - 2 \gm_1 
{w \over z} \over \sqrt {\Dl}} } & \mbox{for $\Dl > 0$}
\end{array}
\right . 
\eqno (4.8)
$$
where $c$ is an arbitrary integration constant.\\

{\bf (b)} $\bt_1 + \gm_2 = 0$ and $\Dl = 0$ with $\bt_2 \gm_1 < 0$: 
In this case we may proceed to derive the explicit time-dependence 
of the Hubble parameters and the corresponding scale functions for 
both subspaces. From Eq.(4.8) we obtain
$$
H_2 = c_1 H_1 + c_2 \eqno (4.9)
$$
where
$$
c_1 = \sqrt {\vert {\gm_1 \over \bt_2} \vert} \; \; , \; \; \; 
c_2 = 1 + h_2 - c_1 h_1 \eqno (4.10)
$$
Now, Eq.(4.9) is inserted into Eq.(4.2a) to give
$$
{\dot H}_1 = {\dl H_1 + \eps \over \zt H_1 + \et} \eqno (4.11)
$$
where the constants $\dl \: , \: \eps \: , \: \zt$ and $ \et$ stand 
for the combinations
$$
\dl = \bt_1 c_1 + \bt_2 \; \; , \; \; \eps = \bt_3 + \bt_1 c_2 
$$
$$
\zt = \al_1 c_1 + \al_2 \; \; , \; \; \et = \al_3 + \al_1 c_2 
\eqno (4.12)
$$
We consider the following cases: \\

{\bf (i)} $\dl \; , \; \zt \; \neq \; 0$: In this case, Eq.(4.11) 
results in
$$
{\zt \over \dl} H_1 + {\zt \over \dl} \left ( {\et \over \zt} 
- {\eps \over \dl} \right ) \ln { \left ( H_1 
+ {\eps \over \dl} \right ) } = t - t_0
\eqno (4.13)
$$
where $t_0$ is an integration constant. Now, Eq.(4.1a) in connection 
with Eq.(4.13), may be easily integrated to give the form of $R(t)$, 
when the condition
$$
{\et \over \zt} - {\eps \over \dl} = 0 \eqno (4.14)
$$
holds. Then, we obtain
$$
\ln {R(t)} \sim A_1 (t - t_0) + {\dl \over 2 \zt} (t - t_0)^2 
\eqno (4.15)
$$
which introduces a quadratic correction to the expected De Sitter 
solution.
\\

{\bf (ii)} $\dl \; , \; \et \; \neq \; 0$ and $\zt = 0$: In this 
case we rediscover the solutions of Ishihara$^{12}$, obtained in 
the GB theory, as a particular case of the general solution. 
Indeed, from Eq.(4.11) we obtain
$$
H_1 = C e^{ {\dl \over \et} t} - {\eps \over \dl} \eqno (4.16)
$$
where $C$ is an arbitrary integration constant. Therefore the 
corresponding external scale function is of the form
$$
\ln {R(t)} \sim \left ( A_1 - {\eps \over \dl} \right ) (t - t_0) + 
C {\et \over \dl} e^{{\dl \over \et} (t - t_0)} 
\eqno (4.17)
$$
Since the external space expands, we must have 
$ A_1 \: > \: {\eps \over \dl}$. For $C = 0$, Eq.(4.17) reads
$$ 
R(t) \sim e^{( A_1 - {\eps \over \dl}) (t - t_0)} \eqno (4.18a)
$$
corresponding again to a De Sitter phase, while for 
$ {\et C \over \dl} \: \ll \: 1$ it yields
$$
R(t) \sim e^{(A_1 - {\eps \over \dl}) (t - t_0)} 
\left ( 1 + C ({\et \over \dl} )^2 
e^{{\dl \over \et} (t - t_0)} \right ) 
\eqno (4.18b)
$$
For $\eps = 0$ Eq.(4.18b) corresponds to the solution of Ishihara 
(Eq.(15) of Ref. [12]) obtained in the framework of the GB theory. 
\\

{\bf (iii)} $\dl = 0 \: , \: \zt \neq 0$: Finally, in this case, 
Eqs. (4.1a) and (4.11) result in
$$
\ln {R(t)} \sim \left ( A_1 - {\et \over \zt} \right ) (t - t_0 ) 
\pm {1 \over 2 \eps \zt^2} 
\left [ \et^2 + 2 \eps \zt (t - t_0) \right ]^{3/2} 
\eqno (4.19)
$$
where, in connection with the numerical results we must have 
$A_1 > {\et \over \zt}$. 

In concluding, we see that the coupling constants $\lm_{(m)}$ 
may not be arbitrary. In every case, they should satisfy certain 
algebraic relations, depending on the form of the corresponding 
solution around the attracting points.

Since both Eqs.(4.2) are almost of the same functional form, in 
all of the preceding cases, similar functional results may be 
obtained for the internal space, through the solution of Eq.(4.2b). 
In this case, however, we must take into account the fact that the 
numerical results indicate that the extra dimensions contract 
$(A_2 < 0)$. This argument may lead to additional constraints on 
the coupling constants $\lm_{(m)}$.

\section*{V. Discussion and Conclusions}

In the present paper we have studied the time evolution of 
anisotropic, ten-dimensional cosmological models in the framework of 
a quartic Lovelock-Lagrangian theory of gravity$^{1,9-11}$. 
The cosmological models under consideration consist of one time 
direction and two homogeneous and isotropic subspaces: A 
three-dimensional {\em external} space, which represents the 
ordinary Universe, and a compact {\em internal} space, which 
is constituted by the extra dimensions. The evolution of the 
Universe depends on four free parameters. These are the 
coefficients $\lm_{(m)}$  which introduce the extra curvature terms 
in the gravitational Lagrangian (m = 0, 2, 3, 4). They are to be 
regarded as the coupling constants$^{3,9}$. Since we have 
considered models of an already compactified internal 
space$^{17}$, we accordingly examine the process of its 
contraction$^{15-17,19}$. 

The Universe is filled with matter in the form of an anisotropic 
perfect fluid. Given an equation of state for the matter content 
of each subspace, the time evolution of the Universe is completely 
determined by a system of three second-order, non-linear 
differential equations, consisting of the field equations $(2.4b)$ 
and $(2.4c)$ together with the conservation law (2.6). The initial 
value field equation $(2.4a)$ corresponds to a {\em constraint} 
which should be satisfied by the cosmological solutions. 
As regards the energy momentum tensor, we have considered two cases: 
{\bf (a)} Vacuum models, $\rh = 0$, in connection with spatially 
flat subspaces and {\bf (b)} Models of an heterotic superstring gas, 
with $p_{ext} = {1 \over 3} \rh$ and $p_{int} = 0$, in connection 
with positively curved subspaces.

The three independent equations may be subsequently expressed in the 
form of a first order system, Eqs.(2.8), involving the Hubble 
parameters $H_{ext} \:, \: H_{int}$ and the corresponding scale 
functions $R(t) \:, \: S(t)$ of the two factor spaces. This system 
is evaluated numerically, for a wide range of initial conditions of 
the form $ H_0^{ext} > 0$ and $H_0^{int} < 0$. Its solutions may be 
represented by curves in the $H_{ext}-H_{int}$ plane. Those curves 
correspond to the {\em "orbits"} of the dynamical system under study 
and each one of them, associated with a different set of initial 
conditions, represents a different type of evolution for the 
Universe. 

In the case of vacuum models with flat subspaces $(k_{ext} = 0 
= k_{int})$, the numerical results indicate that for all values 
of the coupling constants involved and also for a wide range of 
initial conditions, the Universe will always end up to evolve 
according to an {\em extended De Sitter solution}, i.e. an 
exponentially expanding external space, accompanied by an 
exponentially contracting internal one. Indeed, in this case the 
Hubble parameters of both subspaces approach constant values in the 
later stages. We have confirmed that those values actually represent 
the {\em attracting points} of the dynamical system under 
consideration$^{11-13}$. The appearence of {\em attractors} 
in the solution of the cosmological field equations is very 
important, since, if a spacetime is an {\em attractor} for a wide 
range of initial conditions, then it may be realized asymptotically 
in the later stages$^{11,13}$. Those results indicate that the 
existence of the non-linear curvature terms in the gravitational 
action may lead to inflation without the use of any phase 
transition$^{19,27-30,36}$.

Furthermore, we have investigated the evolution of the attractors 
under the variation of the normalized coupling constants $\kp_m = 
\lm_{(m)}/\lm_{(1)}$ $(m = 0 \: , \: 3 \:, \: 4)$. In all cases, 
the attracting points are displaced at higher absolute values of 
$H_{int}$ as $\kp_m$ increases from $0$ to $1$. As regards the 
variation of $\kp_3$ and $\kp_4$, this result has a clear physical 
meaning. 

In determining the cosmological behaviour of the model very close to 
the initial singularity, the leading terms to consider are those 
with the highest power in $({1 \over t})$, i.e. those obtained from 
the highest order terms in the gravitational action$^{5,6}$. 
Therefore, the increase of $\kp_m$ $(m = 3 \: , \: 4)$ corresponds 
to a more accurate study of the earlier stages in the evolution of 
the Universe$^{9}$. Then, from Figs. 3b and 4b, we see that 
at those epochs the internal space contracts at higher rates than 
those of the GB theory. This means that in the later stages of the 
time evolution, where the GB theory holds alone, the absolute value 
of the internal Hubble parameter decreases, in order for the extra 
space to achieve stabilization at a small physical 
size$^{15-17,19,32}$.

In the vacuum case it is possible to derive the analytic dependence 
of the scale functions upon time, by linearizing the field equations 
around the values of $H_{ext}$ and $H_{int}$ at the attracting 
points. The corresponding results indicate that the functional 
form of the analytic solution depends on several algebraic 
conditions between the coupling constants $\lm_{(m)}$. Therefore, 
in a Lovelock-Lagrangian theory of gravity, the coupling constants 
may play an important role in determining the cosmological behaviour 
of the model Universe. Nevertheless, the coefficients of each term 
in the Lagrangian either should be determined experimentaly or 
they should be given by some underlying foundamental theory$^{3}$. 
In this context, we have rediscovered the solutions of 
Ishihara$^{12}$, obtained in the framework of the GB theory, as 
particular solutions of the general quartic case.

The cosmological models with matter in the form of a superstring 
perfect gas, in which both subspaces are possitively curved 
$(k_{ext} = 1 = k_{int})$, can be treated only numerically. 
In this case, the evolution of the GB models depends additionally 
on the exact value of the normalized coupling constant $\kp_2$. 
We have obtained three different types of cosmological behaviour: \\

{\bf (a)} Power-law solutions, with no attracting points, 
when $0.1 \leq \kp_2 < 0.65$. \\

{\bf (b)} A Kasner-type model, when $\kp_2 = 0.65$. \\

{\bf (c)} Extended De Sitter models, with an attracting point, 
when $0.65 < \kp_2 \leq 1$.\\

In all cases, the external space expands, while the internal one 
contracts. The inclusion of the contributions of the third and/or 
the fourth order terms in the field equations, simply amounts in 
a modulation of the above results (e.g. see Fig. 6b).
\\
\\
{\bf Acknowledgements:} The authors would like to express their 
gratitude to Professor J. Demaret for his suggestions and his 
comments on the content of this article. Furthermore, they would 
like to thank Dr. A. Anastasiadis for several helpful advices on 
their computer work. One of us (K. K.) would like to thank the 
Greek State Scholarships Foundation for the financial support 
during this work.

\section*{Appendix A}

The cosmological field equations (2.4$b$) and (2.4$c$) may be 
recast inthe form of a first order system, as follows
$$
{\dot H}_{ext} = G_1 \: \left ( H_{ext}\, , \; H_{int} \, , 
\; X \, , \; Y \right ) 
\eqno (A.1a)
$$
$$
{\dot H}_{int} = G_2 \: \left ( H_{ext}\, , \; H_{int} \, , 
\; X \, , \; Y \right ) 
\eqno (A.1b)
$$
$$
{\dot X} = - \: X \: H_{ext} \eqno (A.1c)
$$
$$
{\dot Y} = - \: Y \: H_{int} \eqno (A.1d)
$$
where
$$
X^2 \; = \; {k_{ext} \over R^2} \; , \; \; 
Y^2 \; = \; {k_{int} \over S^2} 
\eqno (A.2)
$$
The functions $G_1$ and $G_2$ are given by the expressions
$$
G_1 = {[G_{12}(16\pi p_{int}+G_{20})-
G_{22}(16\pi p_{ext}+G_{10})] \over 
[G_{11}G_{22}-G_{12}G_{21}]} 
\eqno (A.3a)
$$
$$
G_2 = {[G_{21}(16\pi p_{ext}+G_{10})-
G_{11}(16\pi p_{int}+G_{20})] \over 
[G_{11}G_{22}-G_{12}G_{21}]} 
\eqno (A.3b)
$$
where we have set 
$$
G_{10} = \lm_0 + 2 \lm_1 B_{10} + 24 \lm_2 B_{20} 
+ 720 \lm_3 B_{30} + 5760 \lm_4 B_{40} 
\eqno (A.4a)
$$
$$
G_{11} = 4 \lm_1 + 24 \lm_2 B_{21} + 720 \lm_3 B_{31} 
+ 5760 \lm_4 B_{41} 
\eqno (A.4b)
$$
$$
G_{12} = 12 \lm_1 + 24 \lm_2 B_{22} + 720 \lm_3 B_{32} 
+ 5760 \lm_4 B_{42} 
\eqno (A.4c)
$$
$$
G_{20} = \lm_0 + 2 \lm_1 C_{10} + 24 \lm_2 C_{20} 
+ 720 \lm_3 C_{30} + 5760 \lm_4 C_{40}
\eqno (A.4d)
$$
$$
G_{21} = 6 \lm_1 + 24 \lm_2 C_{21} + 720 \lm_3 C_{31} 
+ 5760 \lm_4 C_{41} 
\eqno (A.4e)
$$
$$
G_{22} = 10 \lm_1 + 24 \lm_2 C_{22} + 720 \lm_3 C_{32} 
+ 5760 \lm_4 C_{42} 
\eqno (A.4f)
$$
and the quantities $B_{ij}$ and $C_{ij}$ are given by
\\
$$
B_{10} \; \; = \; \; 3 H_{ext}^2 + 21 H_{int}^2 + 
12 H_{ext} H_{int} + X^2 + 15 Y^2 
\eqno (A.5a)
$$
\bea
B_{20} & = & 35 H_{int}^4 + 27 H_{ext}^2 H_{int}^2 
+ 60 H_{ext} H_{int}^3 + 
4 H_{ext}^3 H_{int} \nn \\ 
& + & 7 H_{int}^2 X^2 + 5 Y^2 \left ( 3Y^2 + X^2 \right ) \nn 
\eea
$$
\hspace{-.4in}+ \; \; \; Y^2 (50 H_{int}^2 + 15 H_{ext}^2 
+ 40 H_{ext} H_{int}) 
\eqno (A.5b)
$$
\bea
B_{30} & = & 7 H_{int}^6 + 33 H_{ext}^2 H_{int}^4 
+ 8 H_{ext}^3 H_{int}^3 
+ 36 H_{ext} H_{int}^5 \nn \\
& + & 9 Y^4 \left ( H_{ext}^2 + H_{int}^2 \right ) 
+ 3 X^2 Y^4 + Y^6 \nn \\
& + & Y^2 \left ( 15 H_{int}^4 + 34 H_{ext}^2 H_{int}^2 
+ 48 H_{ext} H_{int}^3 + 8 H_{ext}^3 H_{int} \right ) \nn 
\eea
$$
\hspace{-.4in}+ \; \; \; 10 H_{int}^2 X^2 Y^2 + 7 H_{int}^4 X^2 
+ 12 H_{ext} H_{int} Y^4  
\eqno (A.5c) 
$$
\bea
B_{40} & = & 39 H_{ext}^2 H_{int}^6 + 12 H_{ext} H_{int}^7 
+ 12 H_{ext}^3 H_{int}^5 \nn \\
& + & 3 H_{ext}^2 Y^6 + X^2 Y^6 \nn \\ 
& + & Y^4 \left ( 21 H_{ext}^2 H_{int}^2 + 12 H_{ext} H_{int}^3 
+ 12 H_{ext}^3 H_{int} \right ) \nn \\
& + & Y^2 \left ( 57 H_{ext}^2 H_{int}^4 + 24 H_{ext} H_{int}^5 
+ 24 H_{ext}^3 H_{int}^3 \right ) \nn
\eea
$$
+ \; \; 9 H_{int}^2 X^2 Y^4 + 15 H_{int}^4 X^2 Y^2 
+ 7 H_{int}^6 X^2 \; 
\eqno (A.5d)
$$
\\
$$
B_{21} \; \; = \; \; 10 H_{int}^2 + 4 H_{ext} H_{int} + 10 Y^2 
\eqno (A.5e)
$$
\\
$$
B_{31} \; \; = \; \; 6 H_{int}^4 + 8 H_{ext} H_{int}^3 + 6 Y^4  
$$
$$
\hspace{.4in}+ \; \; Y^2 \left ( 12 H_{int}^2 
+ 8 H_{ext} H_{int} \right ) 
\eqno (A.5f) 
$$
\\
$$
B_{41} \; \; = \; \; 2 H_{int}^6 + 12 H_{ext} H_{int}^5 
$$
$$
\hspace{.4in}+ \; \;  Y^4 \left ( 6 H_{int}^2 
+ 12 H_{ext} H_{int} \right )
$$
$$
\hspace{.65in}+ \; \; Y^2 \left ( 6 H_{int}^4 
+ 24 H_{ext} H_{int}^3 \right ) + 2 Y^6 
\eqno (A.5g)
$$
\\
$$
B_{22} \; \; = \; \; 20 H_{int}^2 + 2 H_{ext}^2 
+ 20 H_{ext} H_{int} 
$$
$$
\hspace{-.6in}+ \; \; 2 X^2 + 20 Y^2 \eqno (A.5h)
$$
\bea
B_{32} & = & 6 H_{int}^4 + 12 H_{ext}^2 H_{int}^2 
+ 24 H_{ext} H_{int}^3  \nn \\
& + & 4 Y^2 \left ( 3 H_{int}^2 +  H_{ext}^2 
+ 6 H_{ext} H_{int} \right ) \nn
\eea
$$ 
\hspace{-.2in}+ \; \; 4 H_{int}^2 X^2 
+ ( 6 Y^4 + 4 X^2 Y^2) 
\eqno (A.5i)
$$
\bea
B_{42} & = & 12 H_{ext} H_{int}^5 + 30 H_{ext}^2 H_{int}^4 \nn \\
& + & Y^4 \left ( 12 H_{ext} H_{int} + 6 H_{ext}^2 \right ) \nn \\ 
& + & 12 Y^2 \left ( 2 H_{ext} H_{int}^3 
+ 3 H_{ext}^2 H_{int}^2 \right ) \nn
\eea
$$
\hspace{.45in}+ \; \; 12 H_{int}^2 X^2 Y^2 
+ 6 H_{int}^4 X^2 + 6 X^2 Y^4 
\eqno (A.5j)
$$
\\
$$
C_{10} \; \; = \; \; 6 H_{ext}^2 + 15 H_{int}^2 
+ 15 H_{ext} H_{int} + 3 X^2 + 10 Y^2 
\eqno (A.6a)
$$
\bea
C_{20} & = & 15 H_{int}^4 + 45 H_{ext}^2 H_{int}^2 
+ 15 H_{ext}^3 H_{int} 
+ 50 H_{ext} H_{int}^3 + H_{ext}^4 \nn \\ 
& + & 10 Y^2 \left ( 2 H_{int}^2 + 2 H_{ext}^2 
+ 3 H_{ext} H_{int} \right ) \nn
\eea
$$ 
+ \; \; X^2 \left ( 15 H_{int}^2 + 5 H_{ext} H_{int} 
+ H_{ext}^2 \right ) + 5 Y^4 + 10 X^2 Y^2 
\eqno (A.6b)
$$
\bea
C_{30} & = & 26 H_{ext}^3 H_{int}^3 + 36 H_{ext}^2 H_{int}^4 
+ 15 H_{ext} H_{int}^5 + 6 H_{ext}^4 H_{int}^2 \nn \\ 
& + & Y^4 \left ( 6 H_{ext}^2 + 3 H_{ext} H_{int} 
+ 2 H_{int}^2 \right ) \nn \\ 
& + & 2 Y^2 \left ( 15 H_{ext}^2 H_{int}^2 
+ 9 H_{ext} H_{int}^3 
+ 2 H_{int}^4 + 9 H_{ext}^3 H_{int} 
+  H_{ext}^4 \right ) \nn \\ 
& + & 2 X^2 Y^2 \left ( 6 H_{int}^2 
+ 3 H_{ext} H_{int} +  H_{ext}^2 \right ) \nn
\eea
$$
\hspace{-.5in}+ \; \; X^2 \left ( 9 H_{int}^4 
+ 10 H_{ext} H_{int}^3 + 2 H_{ext}^2 H_{int}^2 \right ) 
+ 3 X^2 Y^4  
\eqno (A.6c)
$$
\bea
C_{40} & = & 33 H_{ext}^3 H_{int}^5 + 15 H_{ext}^2 H_{int}^6 
+ 15 H_{ext}^4 H_{int}^4 \nn \\ 
& + & 3 Y^4 \left ( 3 H_{ext}^3 H_{int} +  H_{ext}^2 H_{int}^2 
+ H_{ext}^4 \right ) \nn \\
& + & Y^2 \left ( 34 H_{ext}^3 H_{int}^3 + 18 H_{ext}^2 H_{int}^4 
+ 18 H_{ext}^4 H_{int}^2 \right ) \nn \\ 
& + & 3 X^2 Y^4 \left ( H_{int}^2 +  H_{ext}^2 
+ H_{ext} H_{int} \right ) \nn  \\
& + & 6 X^2 Y^2 \left ( H_{ext}^2 H_{int}^2 
+ 3 H_{ext} H_{int}^3 +  H_{ext}^4 \right ) \nn
\eea
$$ 
\hspace{-.3in}+ \; \; 3 X^2 \left ( H_{int}^6 + 5 H_{ext} H_{int}^5 
+  H_{ext}^2 H_{int}^4 \right ) 
\eqno (A.6d)
$$
\\
$$
C_{21} \; \; = \; \; H_{ext}^2 + 10 H_{int}^2 + 10 H_{ext} H_{int} 
+ X^2 + 10 Y^2 \eqno (A.6e)
$$
\bea
C_{31} & = & 3 H_{int}^4 + 6 H_{ext}^2 H_{int}^2 
+ 12 H_{ext} H_{int}^3 \nn \\
& + & 2 Y^2 \left ( 3 H_{int}^2 +  H_{ext}^2 
+ 6 H_{ext} H_{int} \right ) \nn 
\eea
$$
\hspace{-.1in}+ \; \; 2 X^2 H_{int}^2 
+ (3 Y^4 + 2 X^2 Y^2) \eqno (A.6f)
$$
\bea
C_{41} & = & 6 H_{ext} H_{int}^5 + 15 H_{ext}^2 H_{int}^4 \nn \\
& + & 3 Y^4 \left ( 2 H_{ext} H_{int} +  H_{ext}^2 \right ) \nn \\ 
& + & 6 Y^2 \left ( 2 H_{ext} H_{int}^3 
+ 3 H_{ext}^2 H_{int}^2 \right ) \nn 
\eea
$$
\hspace{.45in}+ \; \; 6 H_{int}^2 X^2 Y^2 
+ 3 H_{int}^4 X^2 + 3 X^2 Y^4  
\eqno (A.6g)
$$
\\
$$
C_{22} \; \; = \; \; 5 H_{ext}^2 + 10 H_{int}^2 
+ 20 H_{ext} H_{int} + 5 X^2 + 10 Y^2 
\eqno (A.6h)
$$
\bea
C_{32} & = & 2 H_{int}^4 + 18 H_{ext}^2 H_{int}^2 
+ 4 H_{ext}^3 H_{int} + 12 H_{ext} H_{int}^3 \nn \\ 
& + & 2 Y^2 \left ( 2 H_{int}^2 + 3 H_{ext}^2 
+ 6 H_{ext} H_{int} \right ) \nn
\eea
$$
+ \; \; 2 X^2 \left ( 3 H_{int}^2 + 2 H_{ext} H_{int} \right ) 
+ 2 Y^4 + 6 X^2 Y^2 \eqno (A.6i)
$$
\bea
C_{42} & = & 15 H_{ext}^2 H_{int}^4 + 20 H_{ext}^3 H_{int}^3 + 
3 H_{ext}^2 Y^4 \nn \\ 
& + & 6 Y^2 \left ( 3 H_{ext}^2 H_{int}^2 
+ 2 H_{ext}^3 H_{int} \right ) \nn \\
& + & 3 X^2 \left ( H_{int}^4 + 4 H_{ext} H_{int}^3 \right ) \nn \\
& + & 6 H_{int}^2 X^2 Y^2 \nn
\eea
$$
\hspace{-.4in}+ \; \; 12 H_{ext} H_{int} X^2 Y^2 
+ 3 X^2 Y^4 
\eqno (A.6j)
$$

\section*{Appendix B}

For $\kp_2 = 1$, in the case of vanishing $\kp_3$ and $\kp_4$, the 
evolution of the attracting point $D_2$ with respect to the 
variation of the cosmological constant $\kp_0$, is given by
$$ 
({dH_{ext} \over d \kp_0 }) = ({QQ_1 \over PP}) \eqno (B.1a)
$$
$$
({dH_{int} \over d \kp_0 }) = ({QQ_2 \over PP}) \eqno (B.1b)
$$
The functions $PP$, $QQ_1$, and $QQ_2$ are given by the expressions
$$
PP = F_{11} F_{22} -F_{12} F_{21} \eqno (B.2a)
$$
$$
QQ_1 = F_1 F_{22} -F_2 F_{12} \eqno (B.2b)
$$
$$
QQ_2 = F_{11} F_2 -F_1 F_{21} \eqno (B.2c)
$$
where we have set
$$
F_1 = - [2 + 24 (10 H_{int}^2 - 3 H_{ext}^2 ) ] \eqno (B.3a)
$$
$$
F_2 = - [2 + 24 (H_{ext}^2 + 6 H_{ext} H_{int} )] \eqno (B.3b)
$$
$$
F_{11} = G_{121}G_{20} + G_{12}G_{201} - G_{221}G_{10} 
- G_{22}G_{101} 
\eqno (B.3c)
$$
$$
F_{12} = G_{122}G_{20} + G_{12}G_{202} - G_{222}G_{10} 
- G_{22}G_{102} 
\eqno (B.3d)
$$
$$
F_{21} = G_{211}G_{10} + G_{21}G_{101} - G_{111}G_{20} 
- G_{11}G_{201} 
\eqno (B.3e)
$$
$$
F_{22} = G_{212}G_{10} + G_{21}G_{102} - G_{112}G_{20} 
- G_{11}G_{202} 
\eqno (B.3f)
$$
Now, the quantities $G_{ij}$ are given by
$$
G_{11} = 4 + 24 (B_{21})_{X=Y=0} \eqno (B.4a)
$$
$$
G_{12} = 12 + 24 (B_{22})_{X=Y=0} \eqno (B.4b)
$$
$$
G_{10} = \kp_0 + 2 (B_{10})_{X=Y=0} 
+ 24 (B_{20})_{X=Y=0} 
\eqno (B.4c)
$$
$$
G_{21} = 6 + 24 (C_{21})_{X=Y=0} \eqno (B.4d)
$$
$$
G_{22} = 10 + 24 (C_{22})_{X=Y=0} \eqno (B.4e)
$$
$$
G_{20} = \kp_0 + 2 (C_{10})_{X=Y=0} 
+ 24 (C_{20})_{X=Y=0} 
\eqno (B.4f)
$$
where $(B_{ij})_{X=Y=0}$ and $(C_{ij})_{X=Y=0}$ denote the form of 
the corresponding quantities for $X = 0 = Y$ and the symbols 
$G_{ijk}$ stand for
$$
G_{ijk} =({\partial G_{ij} \over \partial H_k })_{X=Y=0} \eqno (B.5)
$$
in which, $k = 1$ for $H_{ext}$ and $k = 2$ for $H_{int}$.

\section*{References}

\begin{itemize}

\item[1] D. Lovelock, J. Math. Phys. 12, 498 (1971)
\item[2] S. Kobayashi and K. Nomizu, {\em "Foundations of Differential 
Geometry II"}, Wiley Interscience, New York (1969)
\item[3] L. Farina-Busto, Phys. Rev. D 38, 1741 (1988)
\item[4] K. Kleidis and D. B. Papadopoulos, {\em "Cosmological 
solutions in Kaluza-Klein theories of quadratic Lagrangians"}, 
J. Math. Phys., in press (1997)
\item[5] F. M\"uller-Hoissen, Phys. Lett. B 163, 106 (1985)
\item[6] F. M\"uller-Hoissen, Class. Quantum Gravit. 3, 665 (1986)
\item[7] D. Wurmser, Phys. Rev. D 36, 2970 (1987)
\item[8] B. V. Ivanov, Phys. Lett. B 198, 438 (1987)
\item[9] J. Demaret, H. Caprasse, A. Moussiaux, Ph. Tombal, and D.Papadopoulos, 
Phys. Rev. D 41, 1163 (1990)
\item[10] N. Deruelle and L. Farina-Busto, Phys. Rev. D 41, 1172 
(1990)
\item[11] J. Demaret, Y. De Rop, P. Tombal and A. Moussiaux, 
Gen. Relativ. Gravit. 24, 1169 (1992)
\item[12] H. Ishihara, Phys. Lett. B 179, 217 (1986)
\item[13] S. Capozziello, L. Amendola and F. Occhionero, in {\em 
"Relativistic Astrophysics and Cosmology"} Eds. S. Gottl\"ober, J.P.
M\"ucket, V. M\"uller, Word Scientific Publishing, p. 122 (1992) 
 Potsdam (1992)
\item[14] H. C. Lee, in {\em "Introduction to Kaluza-Klein Theories"}, 
Ed. by H. C. Lee, World Scientific, Singapore (1984)
\item[15] E. W. Kolb, D. Lindley and D. Seckel, Phys. Rev. D 30, 
1025 (1984)
\item[16] T. Applequist, A. Chodos and P. G. O. Freund, {\em "Modern 
Kaluza-Klein Theories"}, Addison - Wesley, New York (1987)
\item[17] U. Bleyer, D. E. Liebscher and A. G. Polnarev, Class. 
Quantum Gravit. 8, 477 (1991)
\item[18] J. Garriga and E. Verdaguer, Phys. Rev. D 39, 1072 (1989)
\item[19] R. B. Abbott, S. Barr and S. D. Ellis, Phys. Rev D 30, 
720 (1984)
\item[20] N. Matsuo, Z. Phys C 36, 289 (1987)
\item[21] E. W. Kolb and M. S. Turner, {\em "The Early Universe"}, 
Addison-Wesley, New York (1989)
\item[22] N. Deruelle, Journ. Geom. Phys. 4, 133 (1987)
\item[23] S. Weinberg {\em "Gravitation and Cosmology"}, Wiley and 
Sons Inc., New York (1972)
\item[24] C. W. Misner, K. S. Thorne and J. A. Wheeler, {\em 
"Gravitation"}, Freeman and Co., San Francisco (1973)
\item[25] N. D. Birrell and P. C. W. Davies, {\em "Quantum Fields in 
Curved Space"}, Cambridge Univ. Press, Cambridge (1982)
\item[26] D. Lorenz-Petzold, Phys. Lett. B 197, 71 (1987)
\item[27] M. Mijics, M. S. Morris and W. Suen, Phys. Rev. D 34, 
2934 (1986)
\item[28] A. Starobinsky, Phys. Lett. B 91, 99 (1980)
\item[29] A. Starobinsky and H.-J. Schmidt, Class. Quantum Gravit. 
4, 695 (1987)
\item[30] V. M\"uller, H.-J. Schmidt and A. Starobinsky, Phys. Lett 
B 202, 198 (1988)
\item[31] K. Kleidis and D. B. Papadopoulos, {\em "On the adiabatic 
expansion of the visible space in a higher-dimensional cosmology"}, 
Gen. Rel. Grav., 29, 275 (1997)
\item[32] A. Chodos and S. Detweiler, Phys. Rev. D 21, 2167 (1980)
\item[33] K. Maeda, Phys. Lett. B 166, 59 (1986)
\item[34] N. Deruelle, {\em "The approach to the cosmological 
singularity in quadratic theories of gravity; the Kasner regime"} 
Preprint. CERN-TH 5297/89
\item[35] I. S. Gradshteyn and I. M. Ryzhik {\em "Tables of 
Integrals, Series and Products"}, Academic Press, N. Y. (1965)
\item[36] K. Maeda, Phys. Rev. D 37, 858 (1988)

\end{itemize}

\newpage

\section*{Figure Captions}

{\bf Fig. 1a:} The evolution of the Hubble 
parameters in the GB theory $(\kp_2 = 1)$, for three different 
sets of initial conditions $(H_0^{ext} , H_0^{int})$: 
$(0.75, -0.25)$ [solid line] $(0.85, -0.15)$ [dashed line] 
and $(0.95, -0.05)$ [squares]. The time coordinate is measured 
in units of $10^4 \: t_{Pl}$. 

{\bf Fig. 1b:} The orbits (in the $H_{ext}-H_{int}$ plane) of the 
dynamical system determined by the cosmological field equations 
for a model with flat subspaces, for three different sets of 
initial conditions. All orbits end at the attracting point 
$D_2 \; (0.8866 \: , \: - 0.1375)$. 

{\bf Fig. 2a:} The change in the location of the 
attractor $D_2$ when a non-zero cosmological constant is included, 
for $\kp_0 = 0.5$. 

{\bf Fig. 2b:} The displacement of the attractor on 
the $H_{ext}-H_{int}$ plane for a wide range of values of the 
cosmological constant (squares). Notice the very good agreement 
with the least square fitting result $H_{int} = -  0.075 H_{ext} 
\: - \: 0.071$.

{\bf Fig. 3a:} The evolution of the Hubble 
parameters in the ${\cal L}_{(3)}$-theory, for $\kp_3 = 0.15$ and 
for three different sets of initial conditions $(H_0^{ext} , 
H_0^{int})$: $(0.75, -0.25)$ [solid line] $(0.85, -0.15)$ [dashed 
line] and $(0.95, -0.05)$ [squares]. The time coordinate is measured 
in units of $10^4 \: t_{Pl}$. 

{\bf Fig. 3b:} The displacement of the 
attractor on the $H_{ext}-H_{int}$ plane for a wide range of values 
of the coupling constant $\kp_3$ (squares). There is a very good 
agreement with the plot of a sixth-order polynomial of the form 
$H_{ext} = p_6 \left ( H_{int} \right )$. 

{\bf Fig. 4a:} The evolution of the external 
Hubble parameter for $H_0^{ext} = 0.85$, when several combinations 
of the non-linear terms are gradually included in the field 
equations. The time coordinate is measured in units of $10^4 t_{Pl}$.

{\bf Fig. 4b:} The displacement of the attractor on the $H_{ext}-H_{int}$ 
plane for a wide range of values of the coupling constant $\kp_4$ 
(squares). Notice the very good agreement with the plot of a 
third-order polynomial (solid line) of the form $H_{int} = 
p_3(H_{ext})$.

{\bf Fig. 5a:} The time-evolution of the positively curved external 
space, for several values of the normalized coupling constant 
$\kp_2$. 

{\bf Fig. 5b:} The corresponding evolution of the positively curved 
internal 
space. In both figures the time coordinate is measured in units of 
$10^4 \: t_{Pl}$. Notice the change in the cosmological behaviour 
of both subspaces when $\kp_2 < 0.65$ and when $\kp_2 > 0.65$.

{\bf Fig. 6a:} The orbits (in the $H_{ext}-H_{int}$ plane) of the 
dynamical system determined by the cosmological field equations for 
a model with positively curved subspaces, for three different sets 
of initial conditions when $\kp_2 > 0.65$. All orbits end at the 
attracting point P. 

{\bf Fig. 6b:} The time-evolution of the Kasner solution 
$R(t) \sim t^{0.55}$ for 
the external space, for several values of the normalized coupling 
constant $\kp_3$. Again, the time coordinate is measured in units 
of $10^4 \: t_{Pl}$. Notice the slight modulation in the 
time-evolution when $ 0 \leq \kp_3 \leq 0.75$.

\end{document}